\documentclass[aps,preprint]{revtex4}%
\usepackage{amsfonts}
\usepackage{amsmath}
\usepackage{amssymb}
\usepackage{graphicx}
\usepackage{natbib}%
\usepackage[caption=false]{subfig}
\providecommand{\U}[1]{\protect\rule{.1in}{.1in}}

\begin{document}
\preprint{ }
\title{Shell formation in short like-charged polyelectrolytes in a harmonic trap}

\author{Sandipan Dutta}
\affiliation{Asia Pacific Center for Theoretical Physics, Pohang, Gyeongbuk, 790-784, Korea}
\author{and Y.S. Jho}
\email{ysjho@apctp.org}
\affiliation{Asia Pacific Center for Theoretical Physics, Pohang, Gyeongbuk, 790-784, Korea}
\affiliation{Department of Physics, Pohang University of Science and Technology, 790-784, Korea}

\begin{abstract}
Inspired by recent experiments and simulations on pattern formation in biomolecules by optical tweezers,
a theoretical description based on the reference interaction site model (RISM) is developed to calculate
the equilibrium density profiles of small polyelectrolytes in an external potential. The formalism is applied
to the specific case of a finite number of Gaussian and rodlike polyelectrolytes trapped in a harmonic potential.
The density profiles of the polyelectrolytes are studied over a range of lengths and numbers of polyelectrolytes
in the trap, and the strength of the trap potential. For smaller polymers we recover the results for point charges.
In the mean field limit the longer polymers, unlike point charges, form a shell at the boundary layer. When the
interpolymer correlations are included, the density profiles of the polymers show sharp shells even at weaker trap
strengths. The implications of these results are discussed.  
\end{abstract}

 \pacs{}

\maketitle

\section{Introduction}

Optical tweezers are excellent tools to trap and manipulate
colloidal particles \cite{weiss1999fluorescence}. Focusing an intense laser beam into a
colloidal solution of nanoparticles \cite{fujii2011fabrication,fu2014einstein,
park2014surface,kim2011self} or polymers \cite{yoshikawa2012single,nabetani2007effects,nie2008patterning}
generates a field gradient which can cause their aggregation.
Due to this capability, it serves as a principal technique
for controlled two- and three-dimensional (2D and 3D)
pattern formations in biomolecules which has applications in
optical sorting of biological systems, cells micromachines, and
manipulation of biopolymers \cite{zhang2009self,PhysRevA.73.031402,
fazal2011optical,xavier2012controlled}. In recent experiments
the polymers have been deposited on a 2D substrate by laser
beams \cite{ito2001optical,yoshikawa2012single}. The formation of microstructures in flexible
biomolecules on metallic nanostructures has provided a mech-
anism for their application in the development of biosensors
\cite{jp305247a,shoji2013permanent}. These biopolymers form ring structures under the
laser radiation forces. Such kinds of pattern formations have
also been observed in trapped liquid crystals 
\cite{murazawa2006laser,brasselet2008statics,jeong2014chiral,jeong2015chiral} and in
point-particle plasmas \cite{drewsen1998large,arp2004dust,
PhysRevLett.96.075001,PhysRevLett.93.165004,Arp2005}. The pattern formations in the
trapped systems are often a result of competing effects of
the repulsive interactions, such as electrostatic or hydrophobic
interactions and the trap potential, causing reversible phase
transitions in polymer gels \cite{juodkazis2000reversible} or the shell structure in
plasmas. Many biomolecules, for instance, the rodlike virus
or liquid crystals or the helical DNA or RNA molecules,
have finite sizes and their geometries play a critical role
in the formation of these patterns \cite{jeong2014chiral,jeong2015chiral,
zhang2009self,PhysRevA.73.031402,fazal2011optical}. It is very
important to understand how the finite-sized particles behave
in the trapping potential of the optical tweezers. While the
theoretical and simulation studies on trapped point-charges
are extensive \cite{wrighton2009theoretical,
bruhn2011theoretical,wrighton2010shell, koulakov1998charging,
bedanov1994ordering,kong2003structural,drocco2003structure,
apolinario2008multiple,schweigert1995spectral}, very few theories exist for finite-sized
charges in traps. The objective of this work is to theoretically
study the distribution of charged polymers in a trap potential
to understand the underlying mechanisms of the structure
formation in charged biomolecules. In many colloidal and
plasma systems, the pattern formation is due to the presence
of some short-ranged attractive forces in the system 
\cite{liu2006simulation,euan2015structural,rice2009structure,
campos2013structural,nelissen2005bubble,liu2008self,evers2013colloids}.
Here we show that the pattern formations can occur even in
the absence of the attractive interactions, primarily due to the
competing effects of the trap and the electrostatic repulsions.

In many optical traps the trapping potential can be well
approximated by a harmonic well 
\cite{florin1998photonic,roichman2008influence}; hence in this work
we specialize to the case of harmonic traps. Wrighton et al.
have developed a theory, based on classical density functional
theory (DFT) and hypernetted chain approximation (HNC),
to study the shell formation in a system of finite number of
point charges in a harmonic trap \cite{wrighton2009theoretical,
bruhn2011theoretical,wrighton2010shell}. They found that
strong correlations are essential to the formation of shells
they successfully predict the location, number, and filling
of the shells. The extension of their theory to polymers is
not straightforward because of the additional orientational
degrees of freedom, constraints of connectivity, and finite
size of the polymers. Additionally, in the case of polymers,
due to their finite sizes, intrapolymer correlations have to be
taken into account together with the interpolymer correlations.
The reference interaction site model (RISM) by Chandler
et al. provides a tool to calculate the density profile of the
polymers in the presence of an external potential and include
both kinds of correlations \cite{chandler1986density1,
chandler1986density2}. In this formalism the
equilibrium density at each site of a polymer is a functional
of the external potential and correlations at that site. This
approach, however, is not very convenient, as a coupled
set of nonlinear equations corresponding to each site needs
to be solved to obtain the density profile at each site. For
uniform polymer systems, Schweizer and Curro 
\cite{curro1987theory,schweizer1987integral,schweizer1994prism} have
developed a theory by averaging over the sites of the polymers,
popularly known as the polymer reference interaction site
model (PRISM). The PRISM theory has been successfully
applied to a variety of polymer systems, including polymer
crystallization, symmetric as well as asymmetric polymer
blends, and block copolymers. In the spirit of the PRISM
formalism, we compute the average equilibrium polymer
density in nonuniform systems by replacing the site quantities
by their corresponding site averages. This vastly reduces the
complexity of the problem of solving matrix equations in the
RISM formalism. As a result of this we obtain a single equation
for the site-averaged density of the polymers as a function of
the site-averaged correlations and external potential.

The outline of the paper is as follows. In Sec. \ref{Sec1}
we phenomenologically derive an integral equation for the
equilibrium site-averaged density of polymer in an external
potential based on the RISM formalism. From this equation we
obtain a closure relation to the PRISM equation similar to the
one obtained by Laria, Wu, and Chandler (LWC) for the pair
correlation functions \cite{laria1991reference}, which is the molecular equivalent
of the HNC equation. In the limit of small polymer length we
recover the HNC equation for the point-particle density. We
apply our formalism to the specific case of finite number of
polyelectrolytes trapped in a harmonic potential. We derive the
density profiles of Gaussian polyelectrolytes in the mean field
approximation in Sec. \ref{Sec2}. The polymer-polymer correlations
are calculated using the LWC and PRISM equations in Sec. \ref{Sec3}.
We go beyond the mean field approximation and obtain the
monomer densities with the full many-body correlations. The
dependence of the correlated densities on the geometry of
the polyelectrolytes and the strength of the trap potential
are worked out. In Sec. \ref{Sec4} we briefly look into the density
profiles of rodlike polyelectrolytes and compare them with
the Gaussian polyelectrolytes to investigate dependence of
the shell formation on the polymer model. We discuss the
limitations of the averaging procedure and the range of validity
of our model in Sec. \ref{Sec5}.

\section{The Formalism}
\label{Sec1}
\begin{figure*}[h]
        \centering
           \subfloat{%
              \includegraphics[height=6.2cm,trim={6cm 1cm 7cm 8cm},clip]{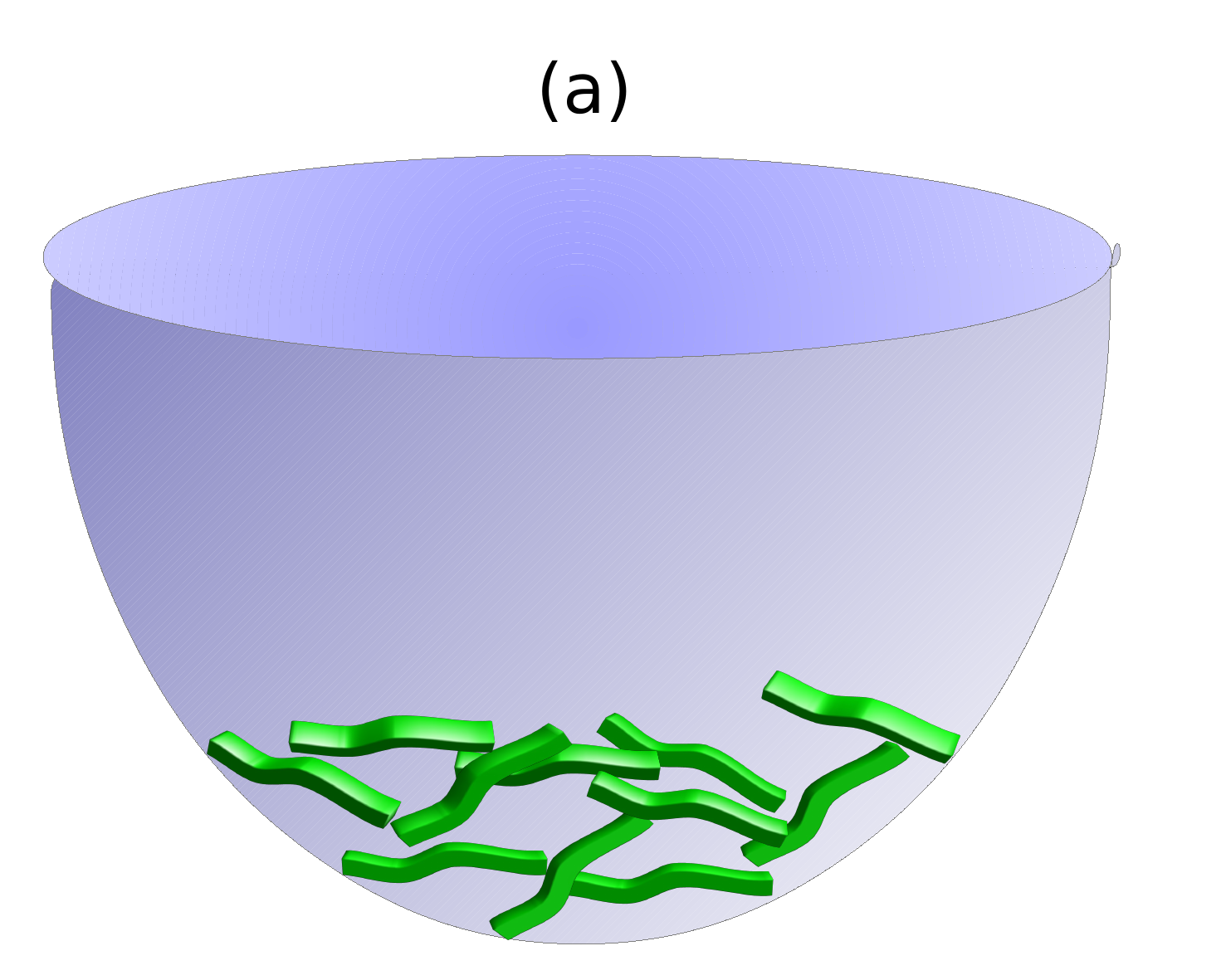}%
           }
           \subfloat{%
              \includegraphics[height=6.2cm,trim={2cm 0 5cm 3cm},clip]{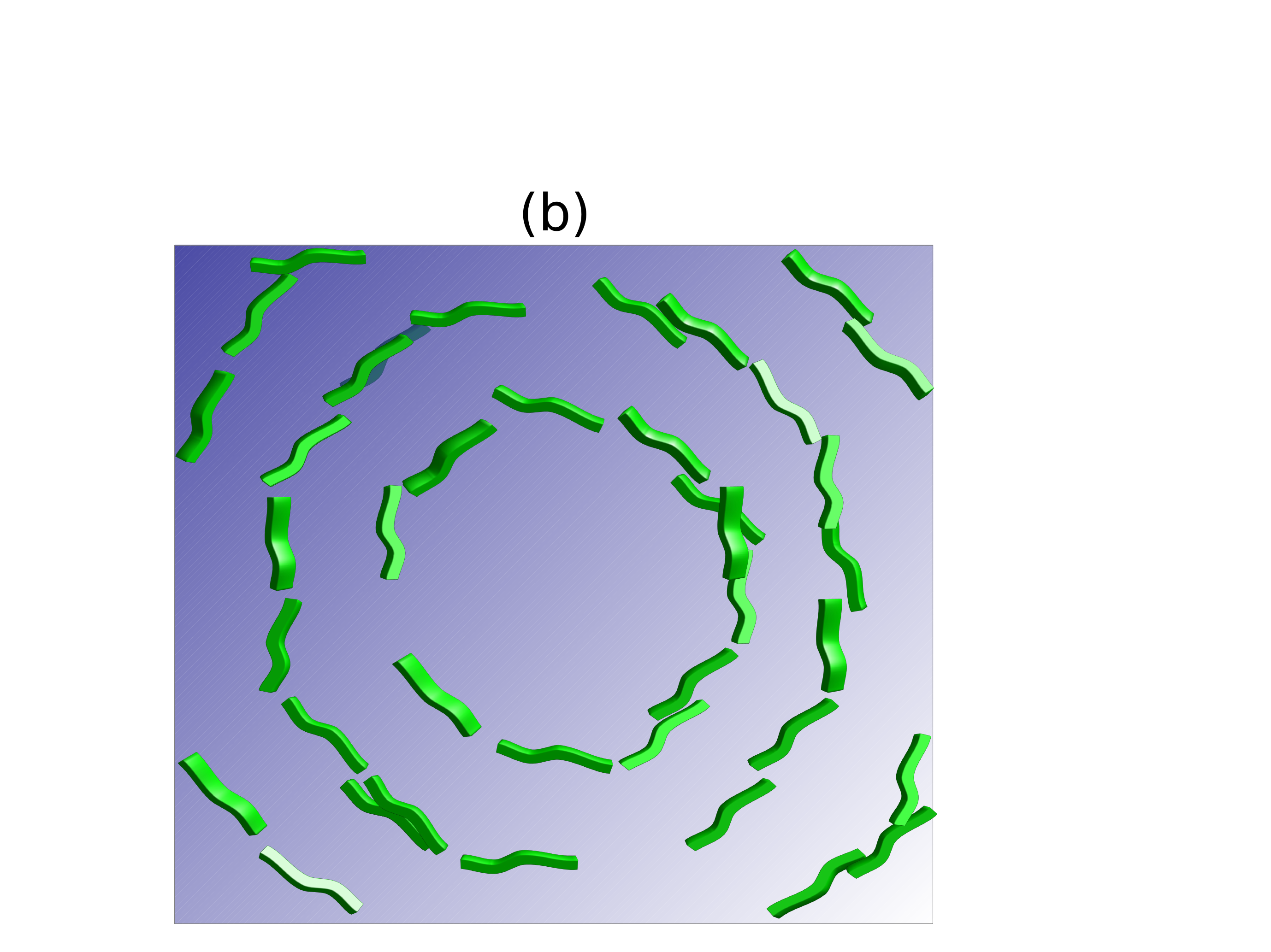}%
           }
           \caption{ (Color online) (a) Schematic diagram showing polyelectrolytes trapped in a harmonic trap. (b) Concentric ring-like structures form due to the competing
           effects of the trap force and the electrostatic forces in 2D. In 3D (not shown) concentric shells are formed.          
            }
           \label{Fig1.1}
 \end{figure*}
Consider a system of $N$ polyelectrolytes each consisting of $L$ monomers. Each monomer has a length $\sigma$ and charge $q$.
For simplicity we assume the hard core diameter of the polymers equals the monomer length $\sigma$.
Thus the length and charge of each polymer would be $L\sigma$ and $Lq$ respectively. The polymers are confined by a harmonic potential of the form
$\phi(\mathbf{r}) = \frac{1}{2}Kr^2$. The schematic diagram of the system is shown in Fig \ref{Fig1.1}.
The Coulomb interaction potential is given by $V(\vert\mathbf{r}-\mathbf{r}^{\prime}\vert) = 1/
\vert\mathbf{r}-\mathbf{r}^{\prime}\vert$. The coordinate of the polymers at the segment $s$ is parameterized by a field $\mathbf{x}(s)$. 
The Hamiltonian of the system reads
\begin{equation}
H = \sum_{i=1}^{N} \int_{0}^{L}ds\phi(\mathbf{x}_i(s)) + \frac{q^2}{2\epsilon}\sum_{i\neq j}\int_0^L ds \int_0^Lds^{\prime}
V(\vert\mathbf{x}_i(s)-\mathbf{x}_j(s')\vert),
\label{eq1.1}
\end{equation}
where $\epsilon$ is the dielectric constant of the medium.
The average inter-monomer distance $r_0$ is related to the average monomer density $\bar{\rho}$ by $\frac{4\pi}{3}r_0^3\bar{\rho} = 1$.
If $R$ is the size of the trap (the position 
of the outermost polymer in the trap), then the average monomer density is given by $\bar{\rho} = \frac{NL}{\frac{4\pi}{3}R^3}$. 
$R$ can be approximately obtained from a force balance condition or finding the position of the outermost polymer such that 
the average forces on it would be zero, $\frac{1}{\epsilon R^2}q^2LN = KR $ \cite{wrighton2009theoretical}.
We define the dimensionless distance by $\mathbf{r}^{\ast} = \mathbf{r}/r_0$ and the dimensionless polymer segment 
field by $\mathbf{x}^{\ast}(s) = \mathbf{x}(s)/r_0$. The dimensionless monomer length is defined
in a similar way, $\sigma^{\ast} = \sigma/r_0$. The dimensionless total potential becomes
\begin{equation}
 \beta V = \frac{\Gamma}{2}\left[\sum_{i=1}^{N}\int_{0}^{L}ds{x_i^{\ast}}^2(s) + \sum_{i\neq j}^{N}
 \int_{0}^{L}ds\int_{0}^{L}d{s^{\prime}}\frac{1}{\vert\mathbf{x}_i^{\ast}(s)
 - \mathbf{x}_j^{\ast}({s^{\prime}})\vert}\right],
 \label{eq1.2}
\end{equation}
where the inverse thermal energy is $\beta = 1/k_BT$ and $\Gamma = \beta q^2/\epsilon r_0$ is the strength of the Coulomb 
interactions among two monomers. The thermodynamic parameter $\Gamma$ measures the strength of the Coulomb potential 
between the monomers relative to the kinetic or thermal energy $k_BT$. For a given trap strength $K$, if some polymers are dropped
into the trap they would come to equilibrium such that the electrostatic repulsions are balanced by the trap potential.
Since $\Gamma$ is determined by the average inter-monomer distance which is obtained from the force balance condition,
$K$ and $\Gamma$ are not independent. In fact they are same in the special case when the distances are scaled with respect to
$r_0$ and the trap is harmonic as we see in equation \eqref{eq1.2}. From now on we use $\Gamma$
for the strength of the trap.

We relate the potential to the density of the polymers through the reference interaction site model developed by Chandler \textit{et al} \cite{chandler1986density2}.
In the rest of the discussions we use only the dimensionless quantities and to keep their notations simple we drop $\ast$. 
The density at site $\alpha$, $\rho_{\alpha}(\mathbf{r})$ can be expressed in terms of the intra-molecular pair correlation function 
$\omega_{\alpha\beta}(\vert\mathbf{r}-\mathbf{r}^{\prime}\vert)$, the local chemical
potential $\psi_{\alpha}(\mathbf{r}) = \mu_{\alpha} - \phi_{\alpha}(\mathbf{r})$ and the direct correlation function 
$c_{\alpha\beta}(\vert\mathbf{r}-\mathbf{r}^{\prime}\vert)$ 
(note we use the direct correlation function of an uniform system for simplicity)
\begin{align}
\rho_{\alpha}(\mathbf{r}) = \prod\limits_{\gamma\neq\alpha}\omega_{\alpha\gamma}*\exp(f_{\gamma}),
 \label{eq1.3}
\end{align}
where 
\begin{equation}
f_{\gamma} = \psi_{\gamma} + \sum_{\eta}c_{\gamma\eta}*\rho_{\eta}.
\label{eq1.35}
\end{equation}
We use the symbol $\ast$ for the convolution operation $p * q = \int d\mathbf{r}^{\prime}p(\mathbf{r})q(\vert\mathbf{r}-\mathbf{r}
^{\prime}\vert)$ and have dropped the position dependence to keep notations simple. Like the PRISM theory \cite{curro1987theory,schweizer1994prism} 
we replace the quantities at each site by the corresponding site averaged quantity. 
This simplifies the algebra of equation \eqref{eq1.3} considerably. 
Summing over the index $\alpha$ and replacing $\omega_{\alpha\gamma}$ by $\omega = \frac{1}{L}\sum_{\alpha\gamma}\omega_{\alpha\gamma}$, we get
\begin{align}
 \rho  & = \sum_{\alpha}\rho_{\alpha} \approx \prod\limits_{\gamma}\omega*\exp(f_{\gamma}).
 \label{eq1.4}
\end{align}
Chandler proposed an additional convolution on RHS of equation \eqref{eq1.3} with the single polymer site-site pair correlations $\omega_{\alpha\beta}$
for polyatomic systems. Here we convolute with the site-averaged pair correlations instead 
\cite{chandler1986density2}
\begin{align}
 \ln\rho & \approx \sum_{\gamma}\ln\left(\omega*\exp(f_{\gamma})*\omega/L\right).
 \label{eq1.5}
\end{align}
Expanding the exponential on RHS of the above equation and keeping till the first order term we get
\begin{align}
 \ln\rho & \approx \sum_{\gamma}\ln\left( 1 + \omega*f_{\gamma}*\omega/L\right) \nonumber\\
 & \approx  \omega*\sum_{\gamma}f_{\gamma}*\omega/L \nonumber \\
 & = \omega*f*\omega/L.
 \label{eq1.6}
\end{align}
In the first step of the derivation we have used of the identity $\int d\mathbf{r}\omega(\mathbf{r}) = 1$.
Using the explicit form of $f$ in equation \eqref{eq1.35} the final expression of the equilibrium density becomes
\begin{equation}
 \ln\rho = \omega*\psi*\omega + \omega*c*\rho*\omega/L,
 \label{eq1.7}
\end{equation}
where $\psi = \sum_{\alpha}\psi_{\alpha}$ and $\rho = \sum_{\alpha}\rho_{\alpha}$. When one of the polymers is fixed at the origin, it would
act as a source of the external potential. In this case $\psi(r) = V(r)$ and the density in equation \eqref{eq1.7}
becomes the pair correlations $\rho(r) = \bar{\rho}g(r)$ \cite{hansen1990theory} 
 
\begin{equation}
 \ln g = \omega*(-\beta V)*\omega + \bar{\rho}\omega*c*(g-1)*\omega.
 \label{eq1.8}
\end{equation}
Using the PRISM equation \cite{schweizer1994prism} 
\begin{equation}
g - 1 = \omega * c * \omega + \bar{\rho}\omega * c * (g - 1),
 \label{eq1.85}
\end{equation}
we see that equation \eqref{eq1.8} is identical to the HNC formalism of Laria, Wu, and Chandler (LWC) \cite{laria1991reference} for molecular systems,
except for an extra convolution of $\omega$ in the second term on the RHS. If we put the distance dependence in equation \eqref{eq1.7} we get the
relation between the monomer density and the external potential
 \begin{align}
  \ln\left(\rho(\mathbf{r})\lambda^3/z\right) = -\int d\mathbf{r}^{\prime}d\mathbf{r}^{\prime\prime}\omega(\vert\mathbf{r}-\mathbf{r}^{\prime}\vert)\beta\phi(\vert\mathbf{r}^{\prime}-\mathbf{r}^{\prime\prime}\vert)
 \omega(r^{\prime\prime}) + \int d\mathbf{r}^{\prime}d\mathbf{r}^{\prime\prime}d\mathbf{r}^{\prime\prime\prime}&\omega(\vert\mathbf{r}-\mathbf{r}^{\prime}\vert)
 c(\vert\mathbf{r}^{\prime}-\mathbf{r}^{\prime\prime}\vert)\times\nonumber\\&\rho(\vert\mathbf{r}^{\prime\prime}-\mathbf{r}^{\prime\prime\prime}\vert)
 \omega(r^{\prime\prime\prime})/L,
 \label{eq1.9}
 \end{align}
where $\lambda = \sqrt{h^2/2\pi mk_BT}$ is the thermal wavelength and $z$ is the fugacity of the system. The direct correlation function
$c(\vert\mathbf{r}-\mathbf{r}^{\prime}\vert)$ in the above equation is calculated using the LWC equation \cite{laria1991reference}
\begin{align}
 \ln g(r) = -\int d\mathbf{r}^{\prime}d\mathbf{r}^{\prime\prime}\omega(\vert\mathbf{r}-\mathbf{r}^{\prime}\vert)\beta V(\vert\mathbf{r}^{\prime}-\mathbf{r}^{\prime\prime}
 \vert)\omega(r^{\prime\prime}) + h(r) -\int d\mathbf{r}^{\prime}d\mathbf{r}^{\prime\prime}\omega(\vert\mathbf{r}-\mathbf{r}^{\prime}\vert)c(\vert\mathbf{r}^{\prime}-\mathbf{r}^{\prime\prime}\vert)
 \omega(r^{\prime\prime}), 
 \label{eq1.10}
\end{align}
and the PRISM equation 
\begin{equation}
 g(r) - 1 = \int d\mathbf{r}^{\prime}d\mathbf{r}^{\prime\prime}\omega(\vert\mathbf{r}-\mathbf{r}^{\prime}\vert)c(\vert\mathbf{r}^{\prime}-\mathbf{r}^{\prime\prime}\vert)
 \omega(r^{\prime\prime}) + \int d\mathbf{r}^{\prime}d\mathbf{r}^{\prime\prime}\omega(\vert\mathbf{r}-\mathbf{r}^{\prime}\vert)c(\vert\mathbf{r}^{\prime}-\mathbf{r}^{\prime\prime}\vert)
 \bar{\rho}h(r^{\prime\prime}),
 \label{eq1.11}
\end{equation}
where $\bar{\rho} = \frac{1}{V_0}\int d\mathbf{r}\rho(\mathbf{r})$ and $h(r) = g(r) -1$. $V_0$ is the volume of the trap. 

We can get rid of the unknown fugacity $z$ on LHS of equation \eqref{eq1.9} by imposing the constraint that there are $N$ polymers
on average in the system 
\begin{equation}
 \int d\mathbf{r}\rho(\mathbf{r}) = NL.
 \label{eq1.12}
\end{equation}
This gives
\begin{equation}
 \rho(\mathbf{r}) = NL\frac{\exp(-\Gamma U(r))}{\int d\mathbf{r}^{\prime}\exp(-\Gamma U(r^{\prime}))},
 \label{eq1.13}
\end{equation}
where
\begin{align}
 U(r,\Gamma,N) & = \int d\mathbf{r}^{\prime}d\mathbf{r}^{\prime\prime}\omega(\vert\mathbf{r}-\mathbf{r}^{\prime}\vert)\phi(\vert\mathbf{r}^{\prime}
 -\mathbf{r}^{\prime\prime}\vert)\omega(r^{\prime\prime}) + \frac{N}{\int d\mathbf{r}^{\prime}\exp(-\Gamma U(r^{\prime}))}\int d\mathbf{r}^{\prime}
 d\mathbf{r}^{\prime\prime}d\mathbf{r}^{\prime\prime\prime}\times\nonumber\\&\omega(\vert\mathbf{r}-\mathbf{r}^{\prime}\vert)
 \bar{c}(\vert\mathbf{r}^{\prime}-\mathbf{r}^{\prime\prime}\vert)\exp(-\Gamma U(\vert\mathbf{r}^{\prime\prime}-\mathbf{r}^{\prime\prime\prime}\vert))
 \omega(r^{\prime\prime\prime}),
 \label{eq1.14}
 \end{align}
with the notation $\bar{c}(\vert\mathbf{r}-\mathbf{r}^{\prime}\vert) = -c(\vert\mathbf{r}-\mathbf{r}^{\prime}\vert)/\Gamma$.

In the rest of the Sections we demonstrate the above formalism by applying it to the case of Gaussian and rod-like polyelectrolytes
in harmonic traps. In the small polymer limit we make connections to the point particle results obtained by Wrighton \textit{et al.}
\cite{wrighton2009theoretical}. We compare our findings with the existing literature on the pattern formation in colloidal systems.

\section{Gaussian polyelectrolytes: Mean field approximation}
\label{Sec2}
In this Section we focus on the Gaussian polyelectrolytes trapped in a harmonic potential with mean field interactions among the polymers.
We calculate their density profiles from equation \eqref{eq1.14} and investigate their dependence on the geometry of the polymers 
as well as the strength of the trap (Coulomb coupling parameter) $\Gamma$ (or the inverse temperature). In the mean field approximation the direct correlation
function in equation \eqref{eq1.14} is replaced by the bare interaction potential $-\Gamma/r$ 
\begin{align}
 U(r,\Gamma,N) & = \frac{1}{2}\int d\mathbf{r}^{\prime}d\mathbf{r}^{\prime\prime}\omega(\vert\mathbf{r}-\mathbf{r}^{\prime}\vert)\vert\mathbf{r}^{\prime}
 -\mathbf{r}^{\prime\prime}\vert^2\omega(r^{\prime\prime}) + \frac{N}{\int d\mathbf{r}^{\prime}\exp(-\Gamma U(r^{\prime}))}\int d\mathbf{r}^{\prime}
d\mathbf{r}^{\prime\prime\prime}\omega(\vert\mathbf{r}-\mathbf{r}^{\prime}\vert) \times\nonumber\\&
 \biggl[\frac{1}{ r^{\prime}}\int_0^{r^{\prime}}dr^{\prime\prime}{r^{\prime\prime}}^2
 \exp(-\Gamma U(\vert\mathbf{r}^{\prime\prime}-\mathbf{r}^{\prime\prime\prime}\vert))\omega(r^{\prime\prime\prime})+
 \int_{r^{\prime}}^Rdr^{\prime\prime}r^{\prime\prime}
 \exp(-\Gamma U(\vert\mathbf{r}^{\prime\prime}-\mathbf{r}^{\prime\prime\prime}\vert))\omega(r^{\prime\prime\prime})\biggr].
 \label{eq2.1}
 \end{align}
In the limit of point particles, $\omega(\vert\mathbf{r}-\mathbf{r}^{\prime}\vert) = \delta(\vert\mathbf{r}-\mathbf{r}^{\prime}\vert)$, we recover the
point particle mean field equation of Wrighton \textit{et al.} \cite{wrighton2009theoretical}.

\begin{figure*}[h]
        \centering
           \subfloat{%
              \includegraphics[height=6.2cm]{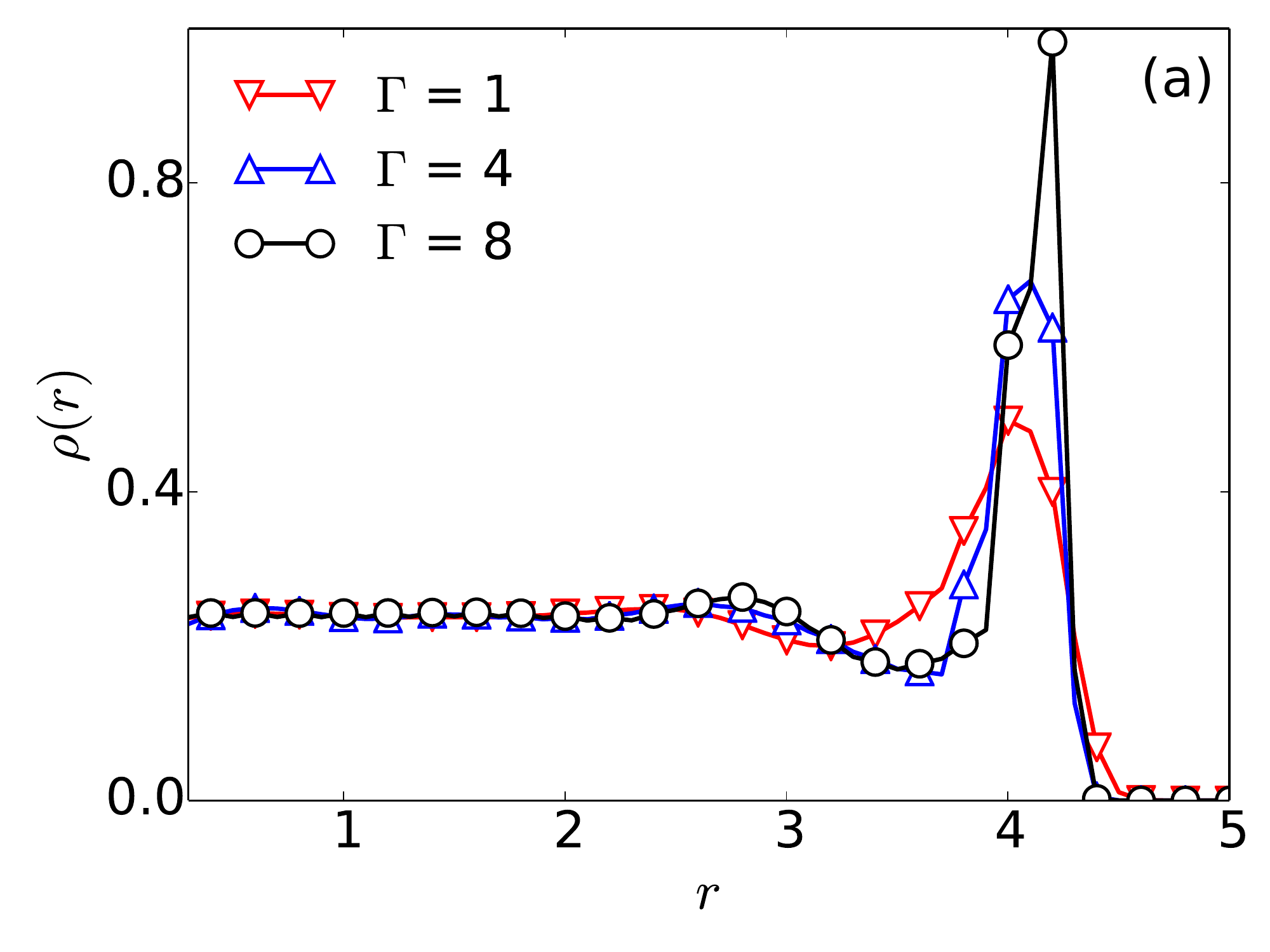}%
           }
           \subfloat{%
              \includegraphics[height=6.2cm]{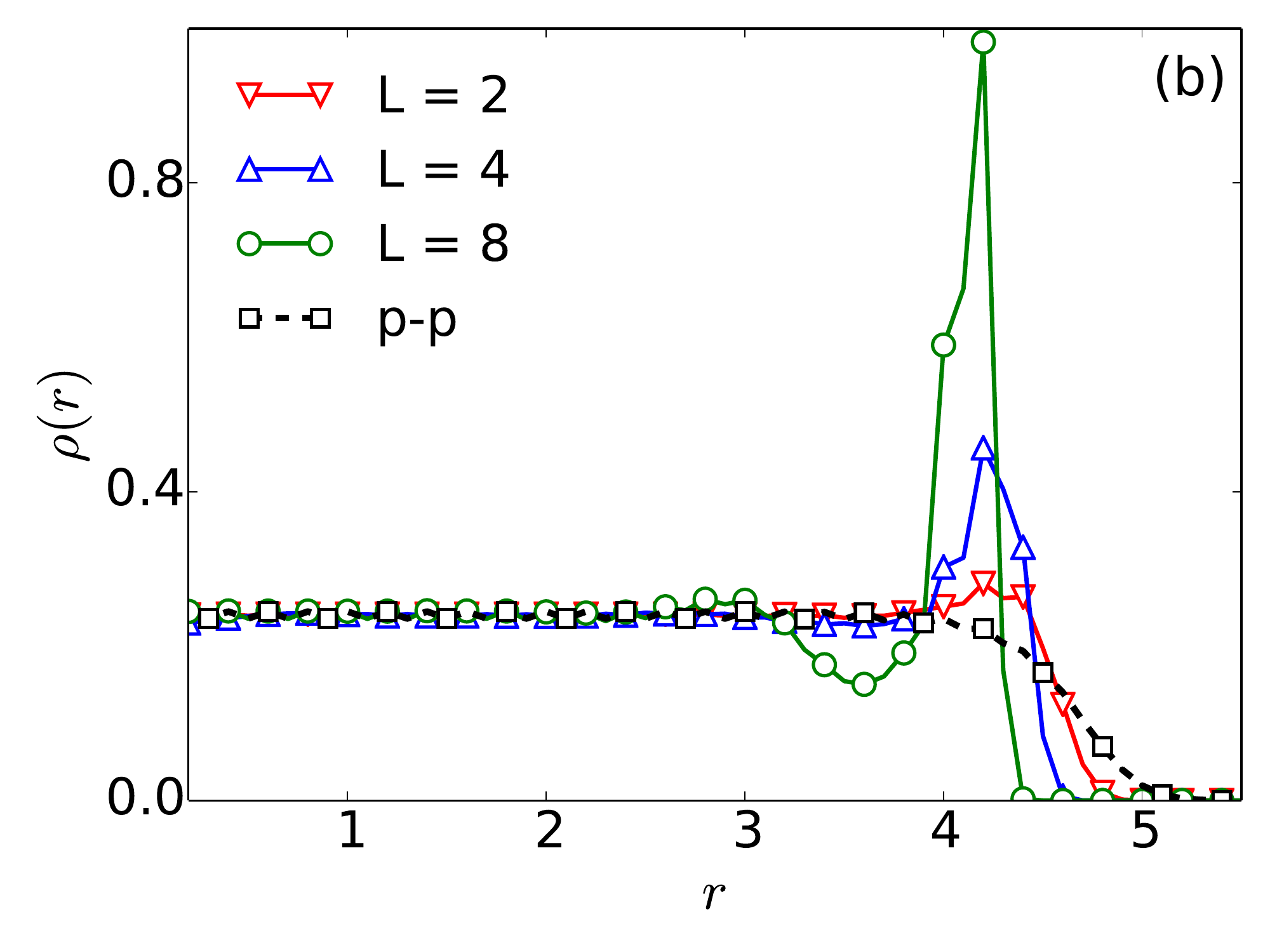}%
           }
           \caption{ (Color online) Mean field: (a) Monomer densities calculated from equation \eqref{eq2.1} for different values of trap strengths $\Gamma$
           for $N = 100$ Gaussian polymers each containing $L = 8$ monomers and monomer length (diameter) $\sigma = 0.5$. 
           Stronger interactions lead to sharper shells. (b) Monomer densities for various lengths $L$ of 
           $100$ polymers with $\sigma = 0.5$ and $\Gamma = 8$. Also shown is the point particle result, p-p.
           Longer polymers have sharper outermost shells.}
           \label{Fig2.1}
 \end{figure*}
\begin{figure*}[h]
        \centering
           \subfloat{%
              \includegraphics[height=6.2cm]{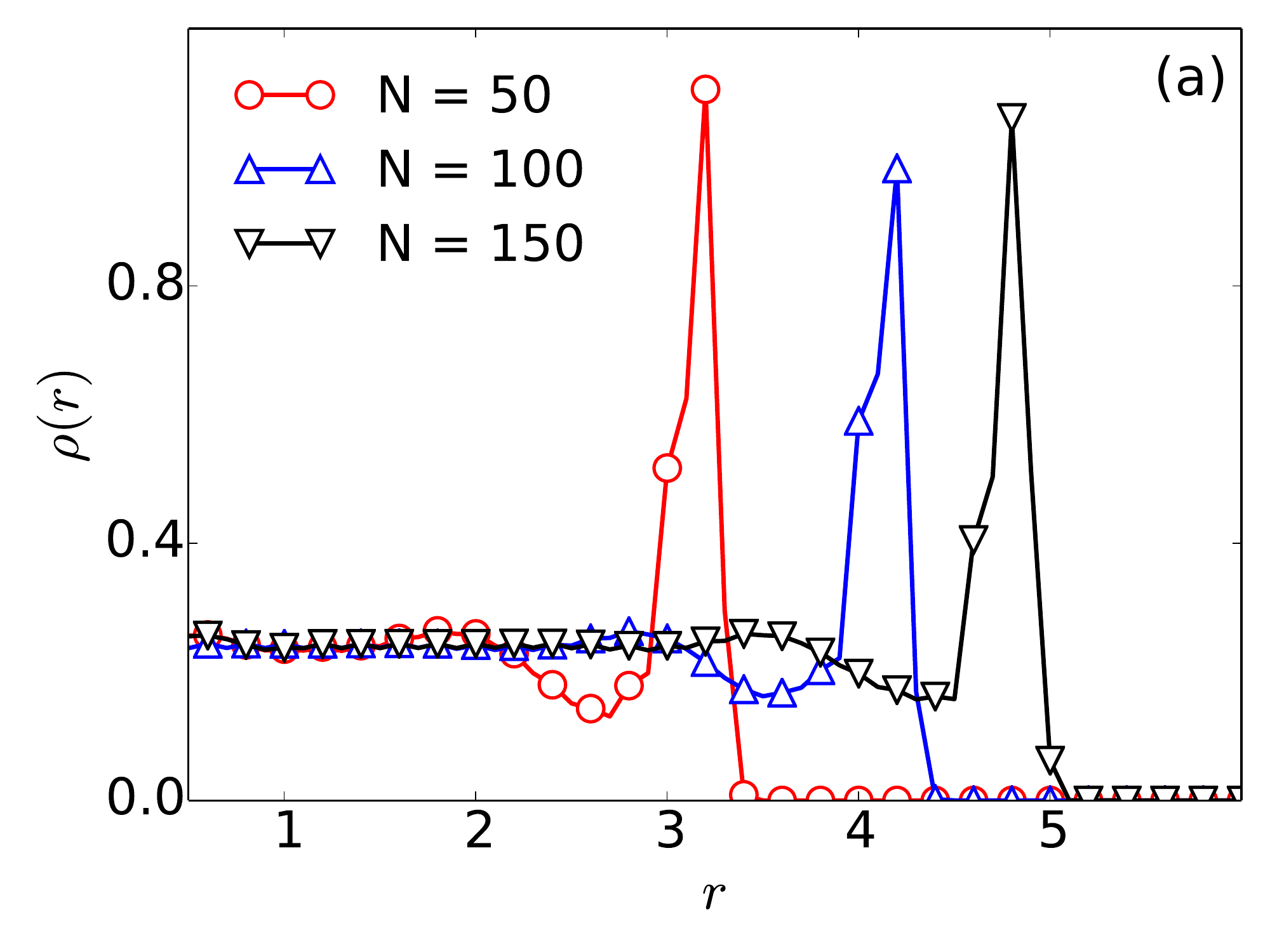}%
           }
           \subfloat{%
              \includegraphics[height=6.2cm]{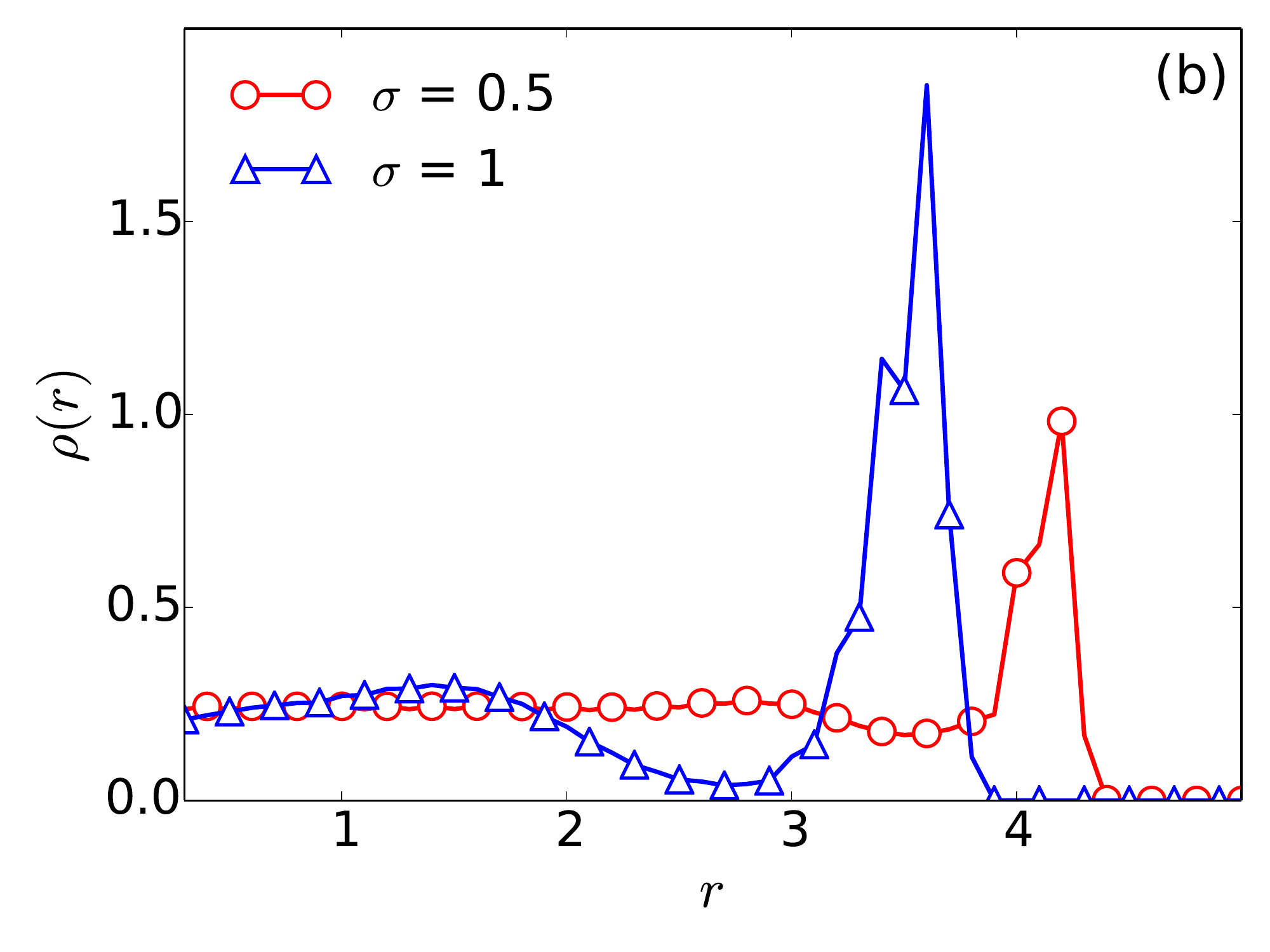}%
           }
           \caption{ (Color online) Mean field: (a) Monomer densities for different values of the number of polymers in the trap $N$ for $L = 8$
           and $\sigma =0.5$ at $\Gamma = 8$. Increasing $N$ does not add any new shell.
           Instead the outermost shell moves outward. (b) Dependence of the monomer density on the monomer length $\sigma$ 
           of the polymers under the same conditions as in (a).}
           \label{Fig2.2}
 \end{figure*}
For Gaussian polymers the single chain structure factor $\hat{\omega}(\mathbf{k})$ in equations \eqref{eq1.14} and \eqref{eq2.1} is given by
\begin{equation}
 \hat{\omega}(\mathbf{k}) = \left(1-f^2-2f/L+2f^{L+1}/L\right)/\left(1-f\right)^2,
 \label{eq2.2}
\end{equation}
where $f = \exp(-k^2\sigma^2/6)$ \cite{schweizer1994prism}. The recursive integral equation \eqref{eq2.1} for $U(r)$ is solved iteratively using 
the Picard's method \cite{hansen1990theory} and using equation  \eqref{eq1.13} we obtain the density. 
In Figure \ref{Fig2.1}-(a) we plot the monomer densities for the polymers of length $L = 8$ for different strengths of the interactions $\Gamma$.
The dimensionless average monomer density is defined as $\bar{\rho}r_0^3 = 3/4\pi = 0.239$. In Figure \ref{Fig2.1} we see that the polymers
close to the center of the trap have a uniform density of $0.239$, while the outermost polymers form a shell which gets sharper with increasing $\Gamma$.
Thus on increasing $\Gamma$ which maybe due to the decrease in 
the temperature of the system or the increase in the polymer charges, the polymers at the boundary would crystallize while the polymers
at the center of the trap would still remain in a fluid state. Though the sharpness of the shells increases no new shells are formed. 
Unlike polymers, the density profile of point-particles is monotonically decreasing and no shells are formed for any value of $\Gamma$. 
The differences between the two cases can be understood from the fact that the 
point particles do not have any internal structure and in the mean field limit we do not get any shells.
For polymers even through the inter-polymer correlations are neglected in the mean field, the stronger fluctuations within the polymer represented by  
$\omega(\mathbf{r})$ in equation \eqref{eq2.1} cause the formation of shells for longer polymers
at couplings $\Gamma \sim 8$ as shown in Figure \ref{Fig2.1}-(b). In the other words the shells appear for the longer polymers
when the Coulomb or trap energy is approximately $8$ times stronger the thermal energy.
The plot clearly shows that for small polymers we recover the point particle limit. Increasing the length of the polymers at a fixed $\Gamma = 8$
makes the outermost shell sharper, hence it is easier for them to crystallize.  
From Figure \ref{Fig2.2}-(a) we see that on increasing the number of polyelectrolytes the outermost shell moves outward. The added polymers move to
the inner fluid layer instead of populating the outermost shell or forming any new shells. Figure \ref{Fig2.2}-(b) depicts that thicker polymers
or polymers with longer monomer lengths move inward because of having lower surface charge density and thus less electrostatic repulsions. 
When either the electrostatic interactions or the trap is strong, the mean field approximation which is valid at weak coupling strength, breaks down. 
In that case the inter-polymer correlations play an important role in their shell structure and can no longer be neglected. 

\section{Gaussian polyelectrolytes: Beyond mean field}
\label{Sec3}

\begin{figure*}[h]
        \centering
           \subfloat{%
              \includegraphics[height=6.2cm]{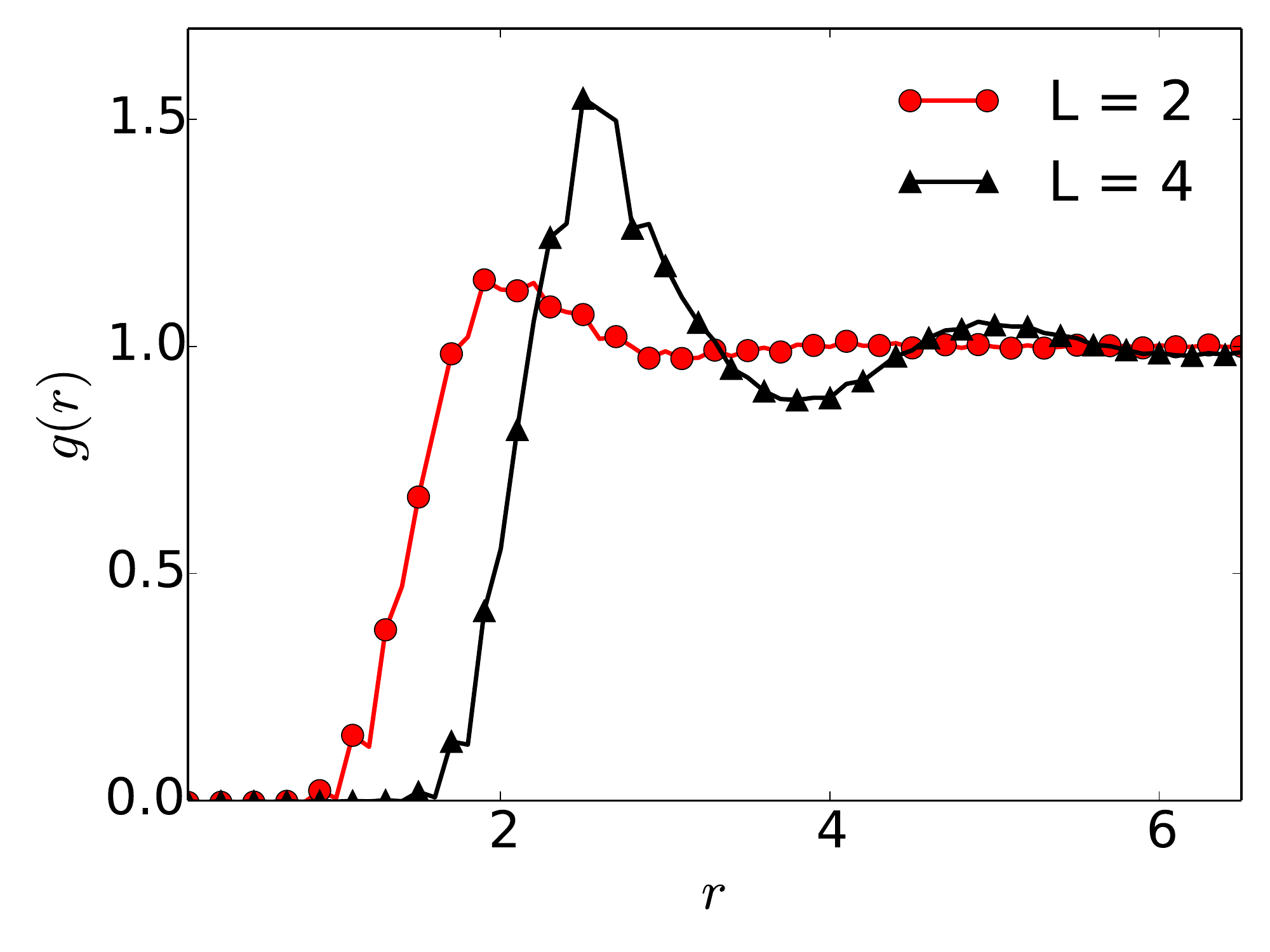}%
           }
           \caption{ (Color online) Inter-polymer pair correlation function for Gaussian polymers of lengths $2$ and $4$ respectively  
           $\Gamma = 4$ and $\sigma = 0.5$. The longer polymer shows peaks in $g(r)$ due to stronger correlations.}
           \label{Fig3.0}
 \end{figure*}

\begin{figure*}[h]
        \centering
           \subfloat{%
              \includegraphics[height=6.2cm]{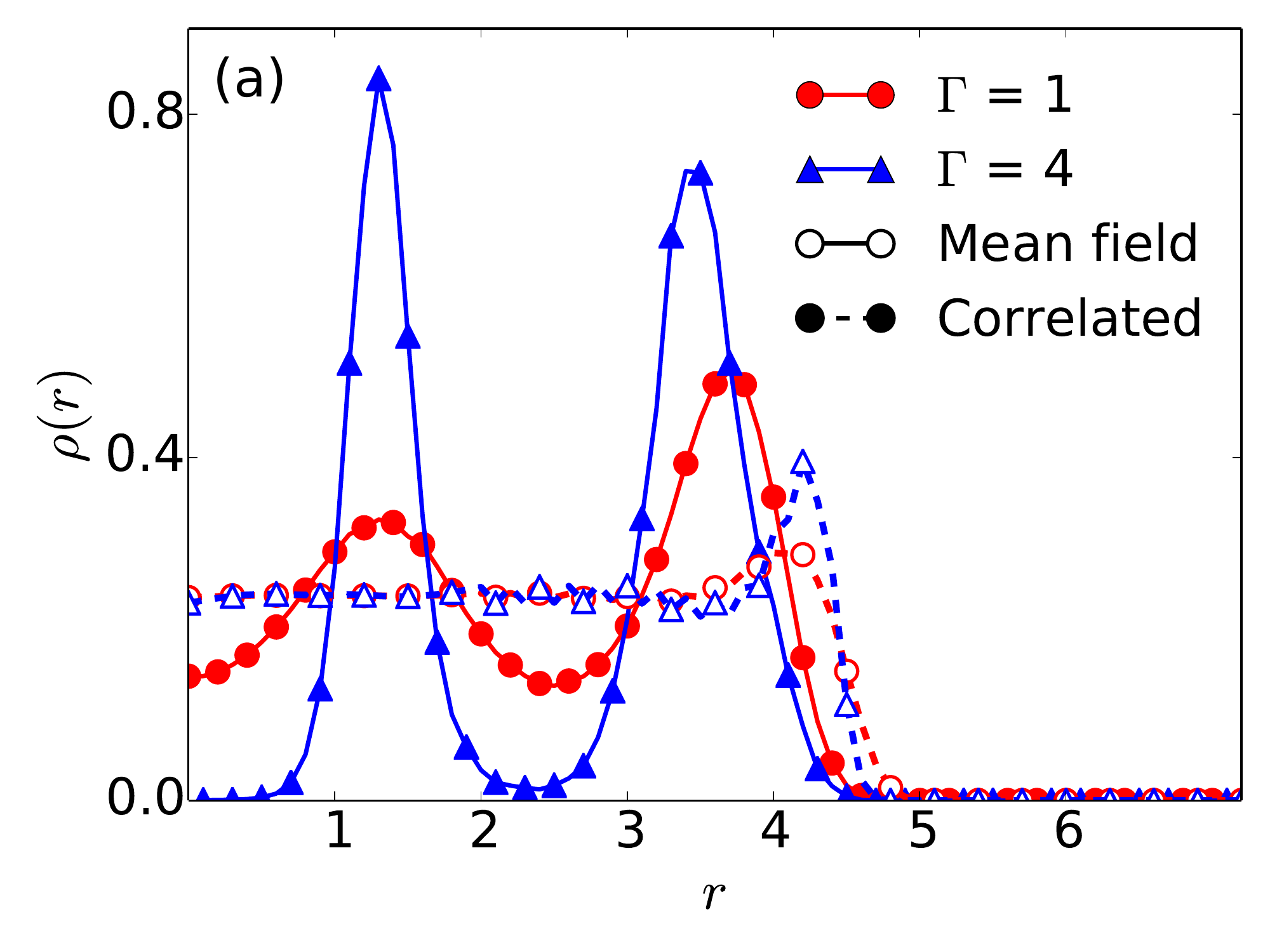}%
           }
           \subfloat{%
              \includegraphics[height=6.2cm]{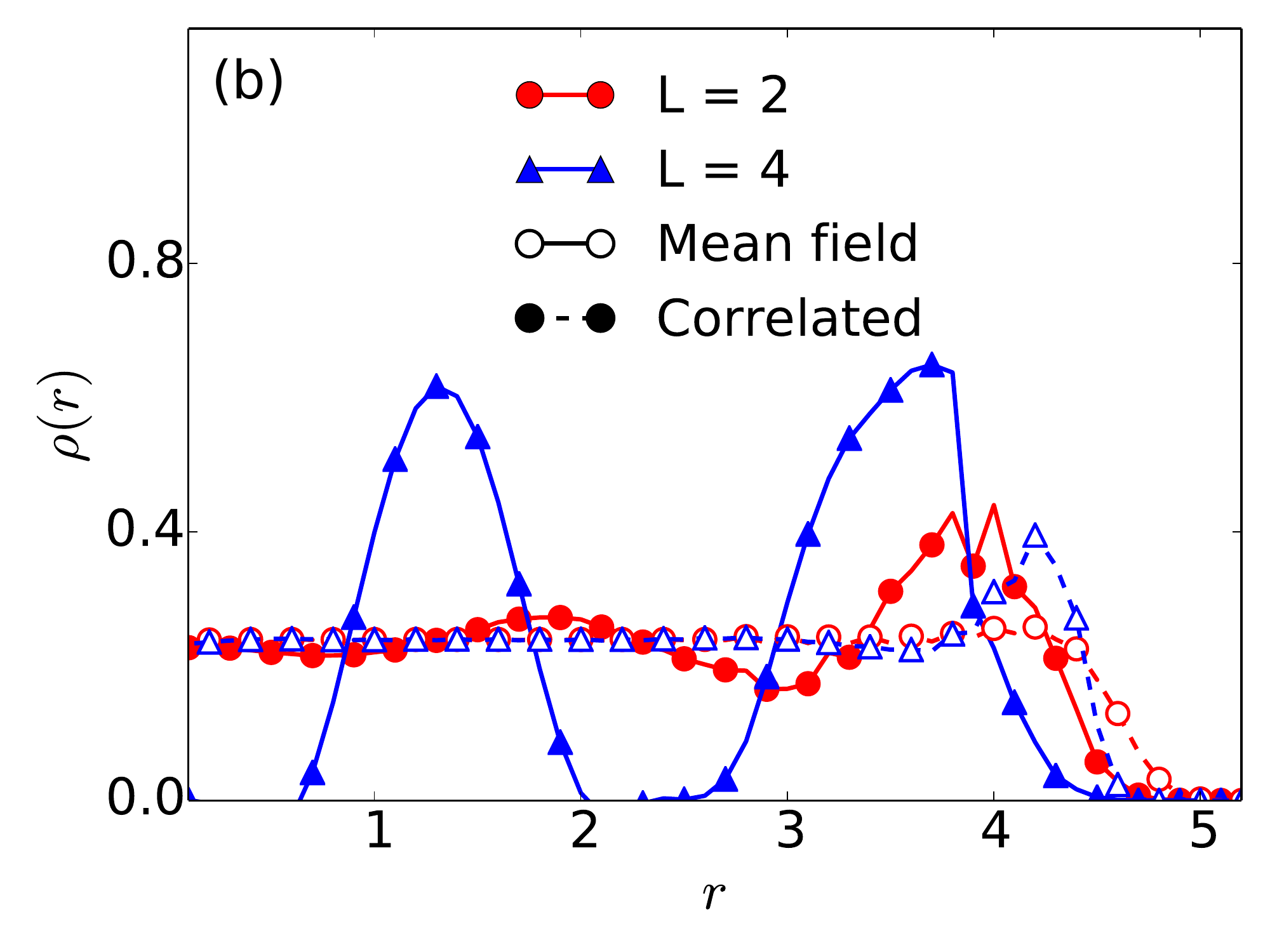}%
           }
           \caption{ (Color online) Correlated densities: (a) The dependence of the correlated density profile (solid, filled) for $100$ polymers 
           with $L = 4$ on $\Gamma$.
           Also shown are the mean field density profiles (dashed, unfilled). Strong correlations at larger $\Gamma$ produce
           sharper shells. (b) Correlated (solid) and mean field (dashed) monomer densities for different lengths $L$ of the polymers. 
           Longer polymers are more strongly correlated and hence have sharper shells. All the polymers have the same $\sigma = 0.5$.}
           \label{Fig3.1}
 \end{figure*}
 \begin{figure*}[h]
        \centering
           \subfloat{%
              \includegraphics[height=6.2cm]{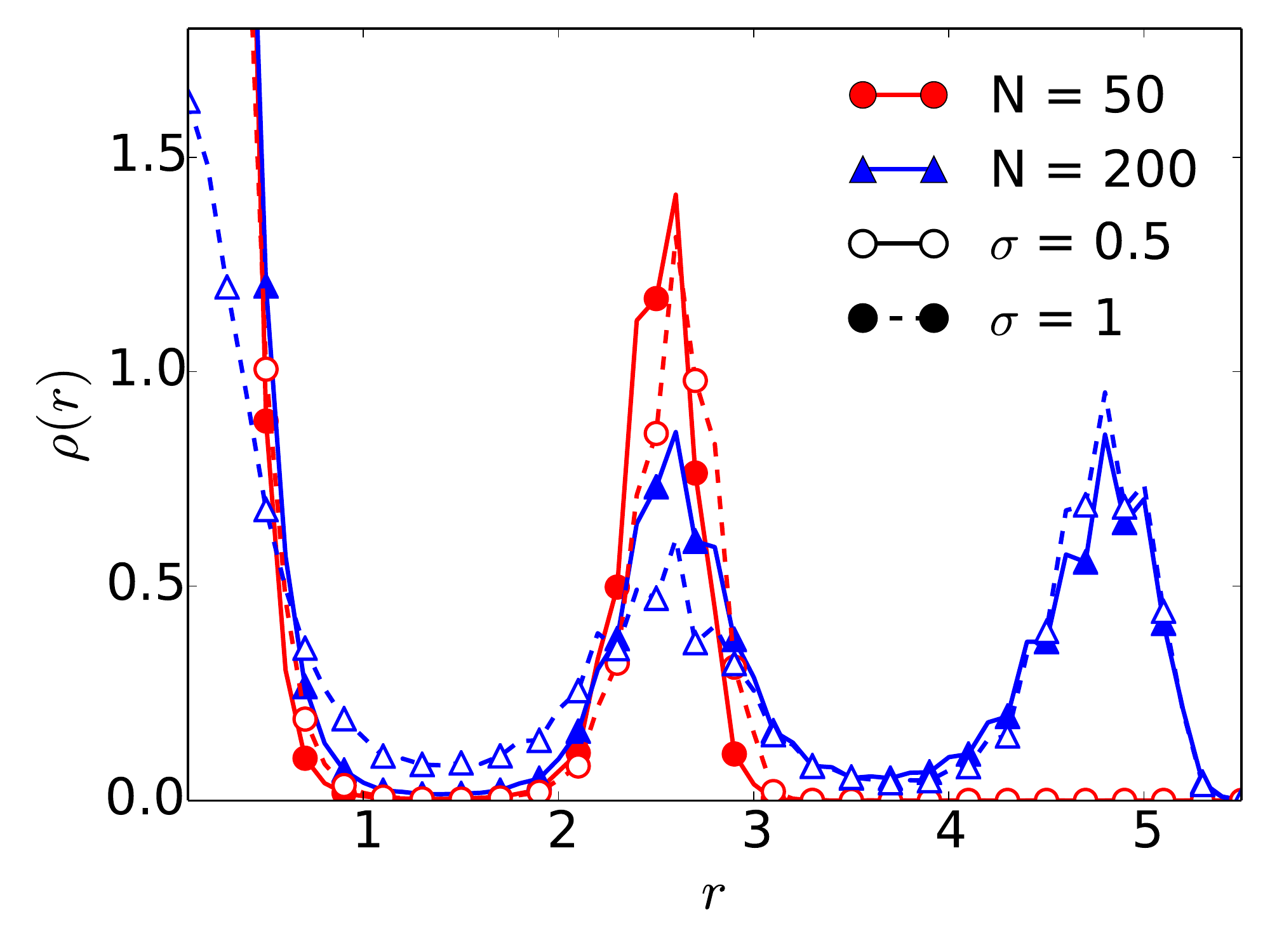}%
           }
           \caption{ (Color online) Correlated densities: On increasing the number of polymers with $L = 4$ at $\Gamma = 4$ in the trap new shells 
           are formed unlike in the mean field case. Polymers of monomer lengths $\sigma = 0.5$ (solid, filled) and $1$ (dashed, unfilled)
           are considered. }
           \label{Fig3.2}
 \end{figure*}

In this Section we explicitly consider the inter-polymer correlations and study their effects on the shell structure of Gaussian polymers.
We solve for the direct correlation function self-consistently from the LWC equation \eqref{eq1.10} and the PRISM equation \eqref{eq1.11} by following
the procedure outlined by Shew and Yethiraj \cite{shew1997integral}. The pair correlation functions in Figure \ref{Fig3.0} clearly portray that
the longer polymers are more strongly correlated as seen from the peaks in the correlation functions. The direct correlation function is then plugged 
into equation \eqref{eq1.14} to obtain the effect potential $U(r)$ and from equation \eqref{eq1.13} the complete density profile. Again Picard's algorithm 
is used to compute $U(r)$ in equation \eqref{eq1.14}. The convergence of the numerical computations becomes increasingly slow for longer polymers and
at large values of $\Gamma$. In that case mixing of different solutions produces faster convergence \cite{hansen1990theory}.  

Figures \ref{Fig3.1}-(a) and \ref{Fig3.1}-(b) show that after taking into account the inter-polymer correlations, sharp shells 
can occur at lower $\Gamma$ or smaller lengths of the polymers. 
In Figure \ref{Fig3.1}-(a) we see that on increasing the trap strength $\Gamma$ the shells become sharper, 
the trend we obtained earlier in the mean field case. In the experiments and simulations on trapped colloidal systems, 
the strength of the trap is the primary controlling parameter. Increasing the strength of the trap leads to the formation of sharper shells
\cite{rice2009structure,euan2015structural,wrighton2009theoretical}.
In this work $\Gamma$ measures the strength of the trap (as well as the Coulomb coupling) and thus our observations from Figure \ref{Fig3.1}-(a) 
qualitatively agrees with these experimental and simulation results. As the shells become sharper and their overlap becomes zero,
it becomes more and more difficult for the polymers to move from one shell to another. Thus the system is effective frozen
in the radial direction but is in a fluid phase within each shell as concluded in References \cite{euan2015structural,wrighton2009theoretical,
bruhn2011theoretical}. At still higher $\Gamma$ the system crystallizes and the liquid state theories are no longer valid.

Figure \ref{Fig3.1}-(b) shows that while the shorter polymers essentially behave like point particles with no shells at moderate $\Gamma$'s, the longer polymers 
by virtue of being more strongly correlated produce sharp shells at such couplings. While in the mean field increasing the number of polymers in the trap does not produce any new 
structure, for the correlated case the behavior is different. New shells appear as the number of polymers in the trap increases as depicted in Figure \ref{Fig3.2}.  
The new shells start forming at the origin and the outermost shell moves outward to accommodate the newer ones similar to the point particle case
\cite{wrighton2009theoretical}. For the point particles 
however the shells start forming at large $\Gamma \ge 10$ values \cite{wrighton2009theoretical}, whereas for longer polymers 
shells form as low as $\Gamma = 2$. Figure \ref{Fig3.2} also shows the dependence of the density structure on the diameter 
(or monomer length) of the polymers. For thicker polymers the sharpness of the shells decreases slightly, although 
the effect of the variation of the polymer diameter is less pronounced after including the correlations.

\begin{figure*}[h]
        \centering
           \subfloat{%
              \includegraphics[height=6.2cm]{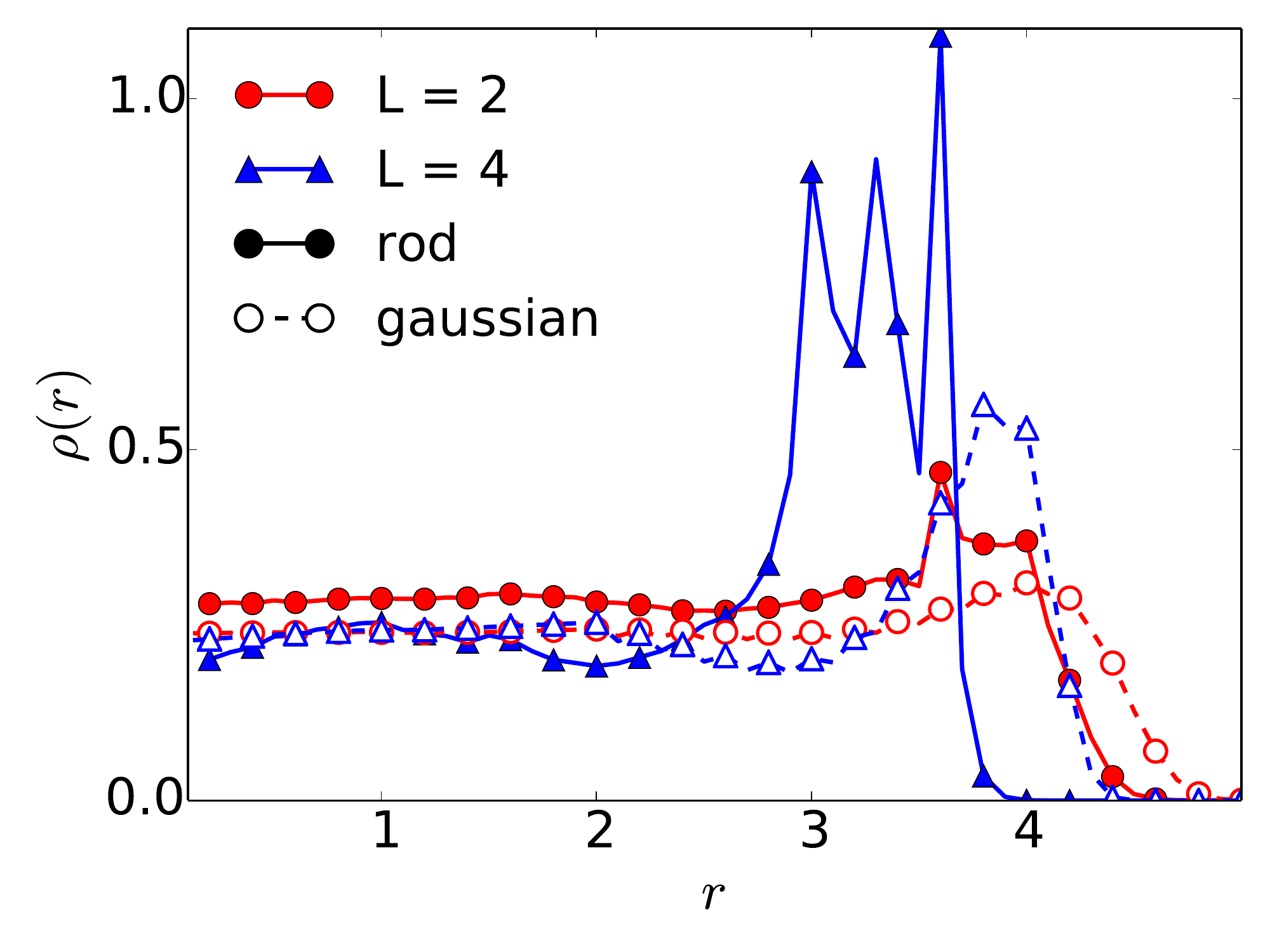}%
           }
           \caption{ (Color online) Mean field: The Monomer densities of $100$ rod-like (solid) and Gaussian polymers (dashed) for different polymer lengths 
            at $\Gamma = 4$. All the polymers have a monomer length of $1$.
           The outermost shell is sharper for the rods than the Gaussian polymers. The three sharp peaks at the boundary are not individual shells
           but part of the outermost shell.}
           \label{Fig4.1}
 \end{figure*}

\section{Rod-like polyelectrolytes}
\label{Sec4}

\begin{figure*}[h]
        \centering
           \subfloat{%
              \includegraphics[height=6.2cm]{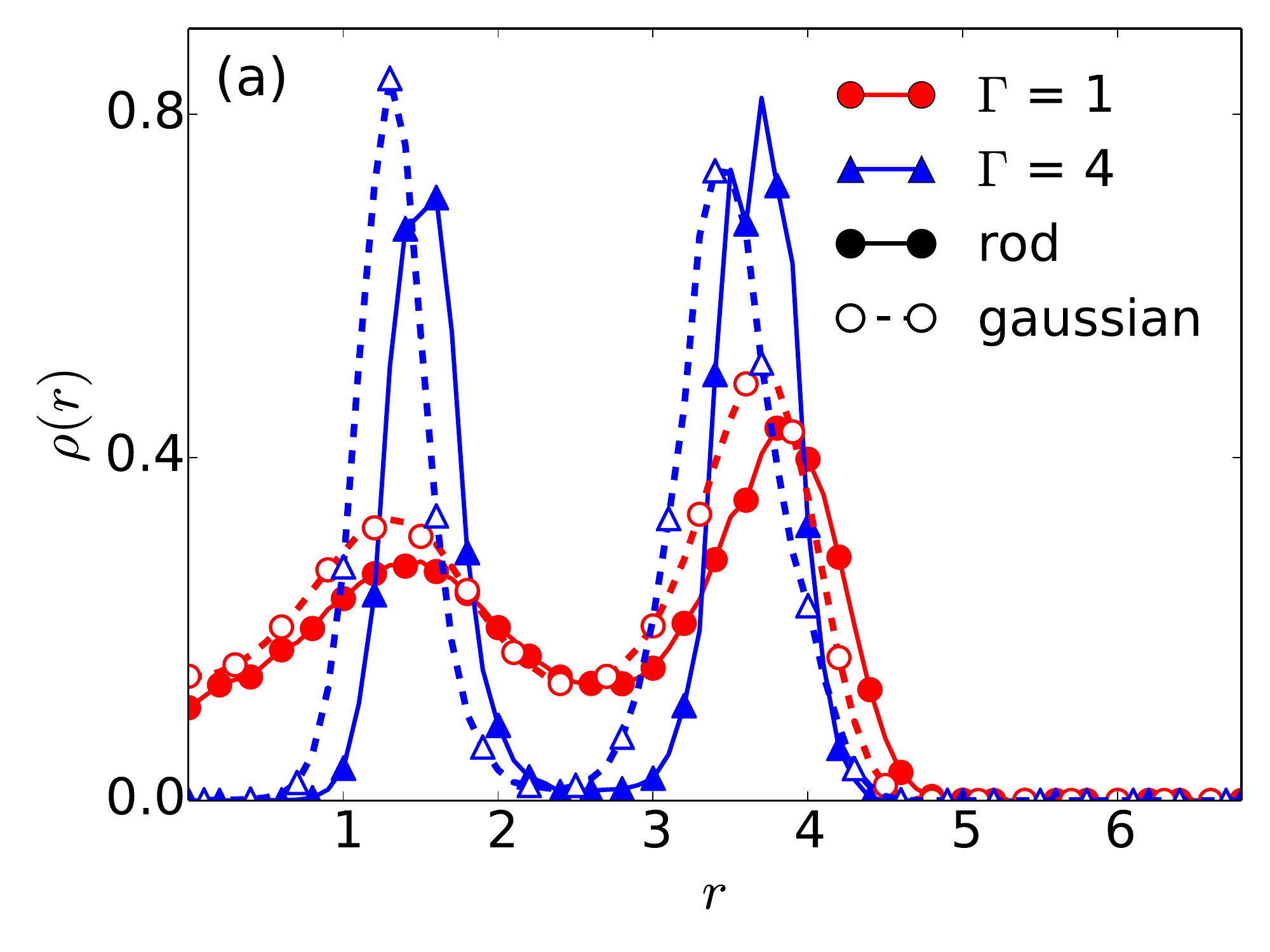}%
           }
           \subfloat{%
              \includegraphics[height=6.2cm]{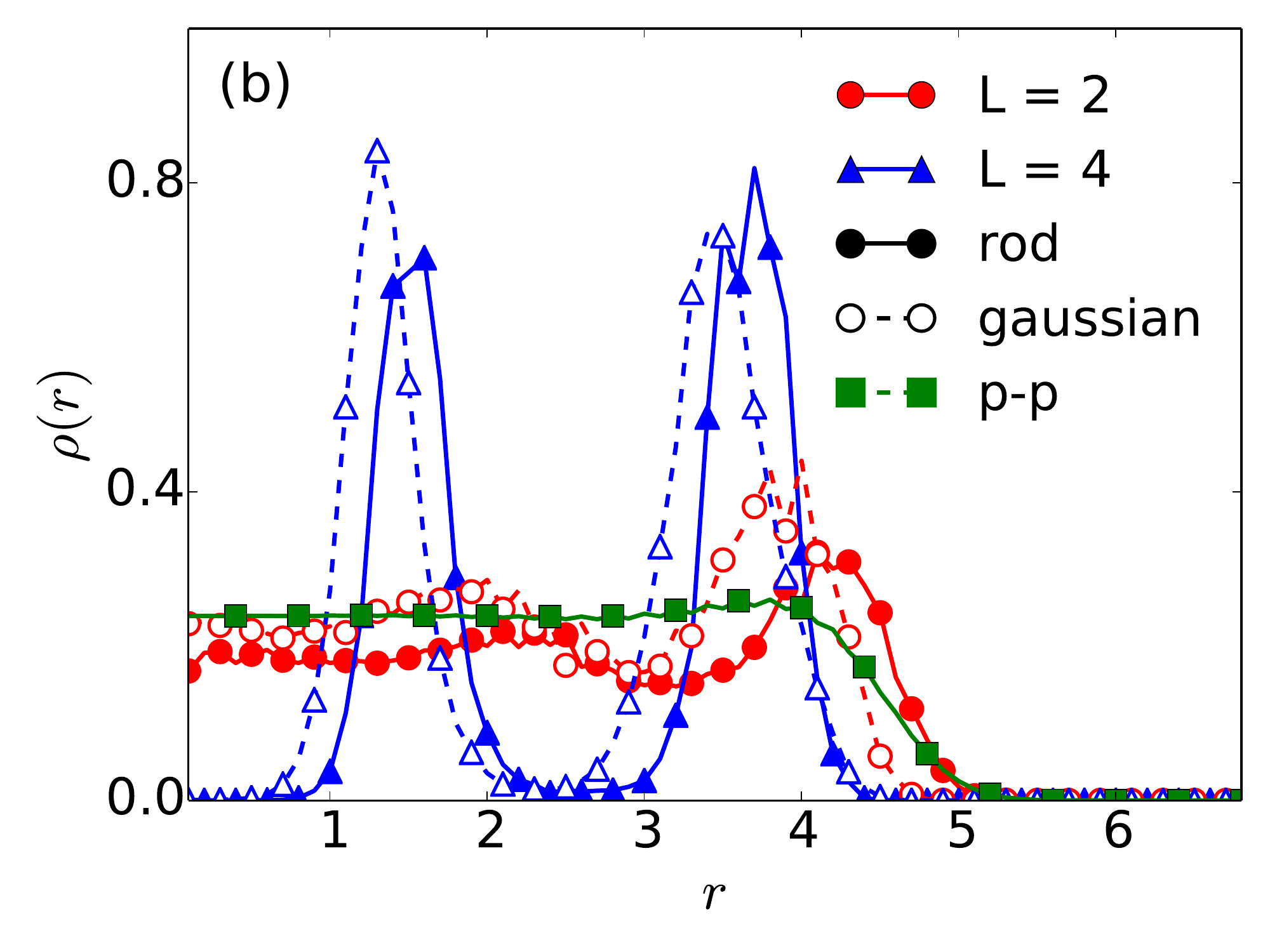}%
           }
           \caption{ (Color online) Correlated densities: The density profile of $100$ rod-like polymers (solid) and Gaussian polymers (dashed) for different values of
           (a) $\Gamma$'s at $L = 4$, and (b) $L$'s at $\Gamma = 4$. The shells moves outward for rod-like 
           polymers due to stronger electrostatic repulsions. All polymers have $\sigma = 0.5$.}
           \label{Fig4.2}
 \end{figure*}
\begin{figure*}[h]
        \centering
           \subfloat{%
              \includegraphics[height=6.2cm]{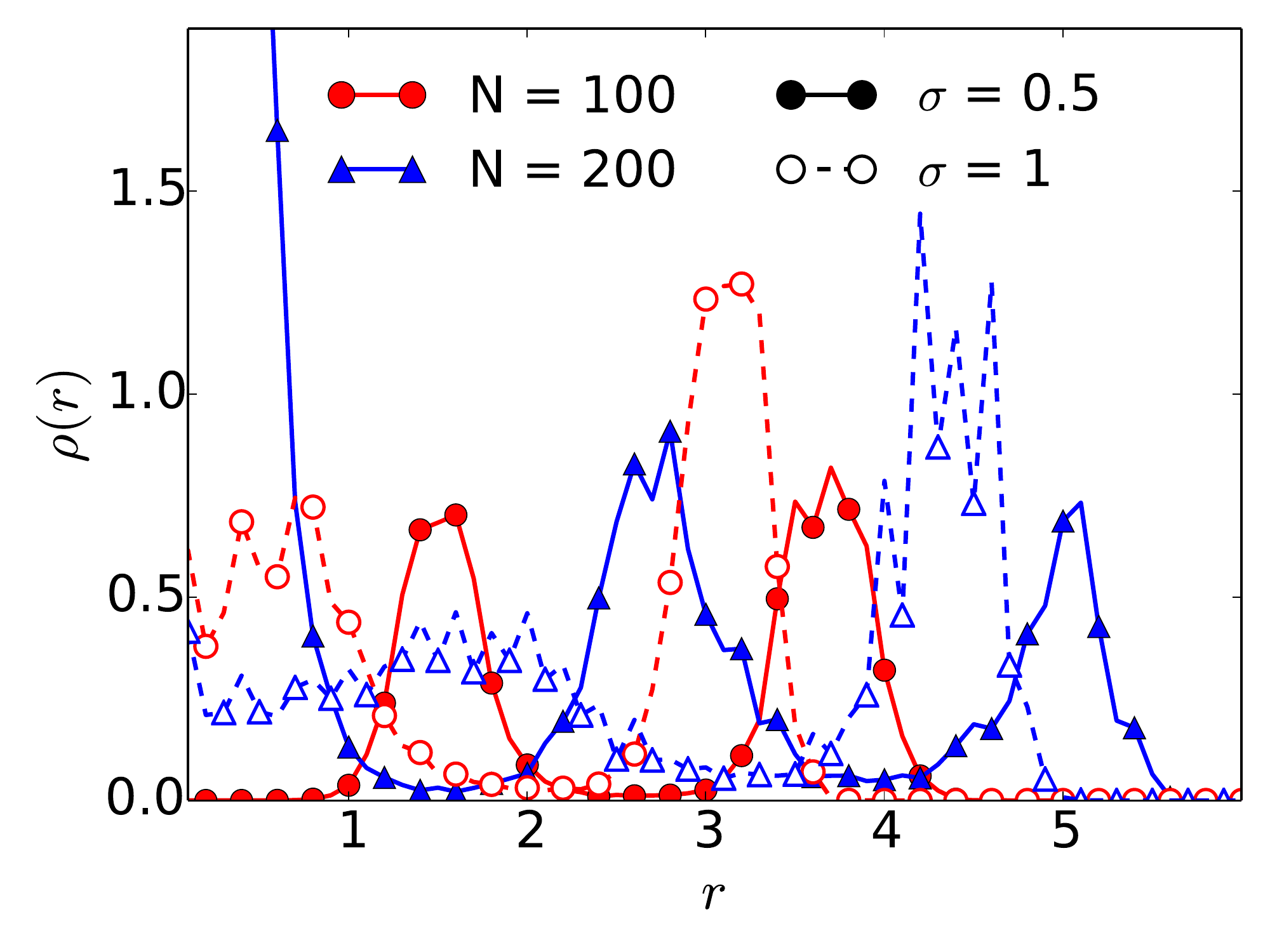}%
           }
           \caption{ (Color online) Correlated densities: The density profile for rod-like polymers of length $4$ at $\Gamma = 4$ showing more 
           shells formed in case of $N = 100$ and $200$ polymers in the trap. Polymers of different monomer lengths $\sigma = 0.5$ (filled)
           and $1$ (unfilled) are also considered.
           The density depends strongly on the diameter of the rods unlike the Gaussian polymers in Figure \ref{Fig3.2}.}
           \label{Fig4.3}
 \end{figure*}

In this Section we look at rigid rod-like polymers which is the opposite limit to the flexibility of the Gaussian polymers we studied 
in the earlier Sections. For rod-like polymers
the single chain structure factor $\hat{\omega}(\mathbf{k})$ in equations \eqref{eq1.14} and \eqref{eq2.1} is given by \cite{shew1997integral}
\begin{equation}
 \hat{\omega}(\mathbf{k}) = 1 + \frac{2}{L}\sum\limits_{j=1}^{L-1}(L-j)\frac{\sin jk\sigma}{jk\sigma}.
 \label{eq4.1}
\end{equation}
The mean field densities for the rods show a sharper outermost shell than the Gaussian polymers in Figure \ref{Fig4.1}. This is due to
the stronger correlations in the rods than the Gaussian polymers which result in their having sharper
outermost shells. Figures \ref{Fig4.2} and \ref{Fig4.3} on the correlated densities show that the shells of rod-like polymers are sharper and are 
shifted outward than the Gaussian polymers. The rigidity causes strong repulsions among the rods compared to the Gaussian polymers and they
move outward to minimize the repulsions. On changing the parameters $\Gamma$ and $L$ in Figures \ref{Fig4.2}-(a) and \ref{Fig4.2}-(b) the rod-like polymers 
qualitatively behave in the same way as Gaussian chains. However the correlated densities of rods in Figure \ref{Fig4.3} have a strong dependence 
on the diameter of the rods (or monomer length) than the Gaussian polymers in Figure \ref{Fig3.2}. This is again due to the rigidity 
of the rods a decrease in the charge density due to the increase in the diameter (or monomer lengths) have a stronger effect on the
correlated densities.

\section{Conclusions and discussions}
\label{Sec5}

We presented a theoretical description of polymer systems in an external potential. We phenomenologically developed our theory for polymer systems 
in an external potential based on the RISM formalism of Chandler \textit{et al.} \cite{chandler1986density1,chandler1986density2} 
and obtained an integral equation for the equilibrium density. Unlike most 
polymer field theoretic approaches where most of the computations are done in mean field, this theory goes beyond mean field and incorporates correlations though the LWC and PRISM
formalisms. We also looked at the specific case of the Gaussian and rod-like polymers trapped in a harmonic potential.
In the small polymer limit we obtain the point particle results of Wrighton \cite{wrighton2009theoretical}.
The density profiles both in the mean field approximation and beyond mean field are explored for different geometries
of the polymers and the strength of the trap potential. This work is effectively a generalization of the formalism developed by
Wrighton \textit{et al.} \cite{wrighton2009theoretical,bruhn2011theoretical} for trapped point charge systems to trapped polyelectrolyte systems. 

This formalism would provide a useful description for the micro-structures that form in polymer colloids confined in optical traps. In recent years
structural transitions in trapped colloids as well as plasmas have attracted the attention of experimentalists as well as theorists
\cite{liu2006simulation,euan2015structural,rice2009structure,campos2013structural,nelissen2005bubble,liu2008self}. The colloidal 
and dusty plasmas have been found to form shell structures in 3D and rings in 2D similar to the predictions by our model. At strong
trap strength we get sharp shells where the inter-shell transitions do not occur while the polymers inside each shell remain in a fluid 
phase as was concluded in Reference \cite{euan2015structural,rice2009structure}. In most of these studies the presence of an attractive
potential or multiple species causes the formation of additional structures. Including attractive interactions in our model would 
enable us to explain the self-assembly of trapped colloids and these new phases. The studies on structural 
transitions of colloidal systems in traps have considered spherical particles and are simpler than the biomolecules for which
we constructed our theory because of their complex geometries and additional length scales. Most experimental studies
focus on trapping of single molecules. Although trapping of multiple charged molecules can done through a technique called
optical bottle \cite{junio2010optical,park2014surface}, the analysis of the pattern formations like the one in this work
have not been done yet, to the best of the authors' knowledge. Simulations and experiments on pattern formations on charged
biomolecules would provide important test for the many-body theories as the one developed here. 

Since the theory is based on the averaging over polymer
sites for inhomogeneous polymer systems, it would describe
the smaller polymers more accurately. For short polymers,
however, the end effects become important and the averaging
process would run into problems. The problem with the
averaging related to the effects of the end points would not
arise in ring polymers. For longer polymers the computations
of the correlations become increasing difficult. The equation
for the density has been derived through linearization, which
would be valid for weak to moderate couplings. Simulations
have to be performed to check the accuracy of the model at
strong coupling. Most real life systems are better described
by semiflexible polymers of which the Gaussian and the
rodlike polymers are special cases. The semiflexible polymers,
however, have an additional directional degree of freedom
which adds to the complexity of the problems. We will tackle
these problems in a subsequent paper.

\section{ACKNOWLEDGMENT}

This research was supported by Basic Science Re-
search Program through the National Research Founda-
tion of Korea (NRF) funded by the Ministry of Edu-
cation (Grants No. NRF2015R1D1A1A09061345 and No.
NRFC1ABA00120110029960).
 
\bibliography{trap}

\end{document}